\documentclass[twocolumn,pra,10pt]{revtex4}
\usepackage{graphicx,amsmath,amssymb,xspace,ntheorem,hyperref}
\usepackage{color}
\usepackage[T1]{fontenc}

\newtheorem*{proof}{Proof}

\begin{document}
%\linenumbers

\newcommand{\ketbra}[2]{| #1\rangle \langle #2|}
\newcommand{\braket}[2]{\langle #1|#2 \rangle}
\newcommand{\ket}[1]{| #1 \rangle}
\newcommand{\bra}[1]{\langle #1 |}
\newcommand{\Tr}{\mathrm{Tr}}
\newcommand\F{\mbox{\bf F}}
\newcommand{\h}{\mathcal{H}}

\newcommand{\PSD}{\textup{PSD}}

\newcommand{\X}{\mathcal{X}}
\newcommand{\Y}{\mathcal{Y}}
\newcommand{\Z}{\mathcal{Z}}
\newcommand{\sspan}{\mathrm{span}}
\newcommand{\kb}[1]{\ket{#1} \bra{#1}}
\newcommand{\pos}{D}

\newcommand{\thmref}[1]{\hyperref[#1]{{Theorem~\ref*{#1}}}}
\newcommand{\lemref}[1]{\hyperref[#1]{{Lemma~\ref*{#1}}}}
\newcommand{\corref}[1]{\hyperref[#1]{{Corollary~\ref*{#1}}}}
\newcommand{\eqnref}[1]{\hyperref[#1]{{Equation~(\ref*{#1})}}}
\newcommand{\claimref}[1]{\hyperref[#1]{{Claim~\ref*{#1}}}}
\newcommand{\remarkref}[1]{\hyperref[#1]{{Remark~\ref*{#1}}}}
\newcommand{\propref}[1]{\hyperref[#1]{{Proposition~\ref*{#1}}}}
\newcommand{\factref}[1]{\hyperref[#1]{{Fact~\ref*{#1}}}}
\newcommand{\defref}[1]{\hyperref[#1]{{Definition~\ref*{#1}}}}
\newcommand{\exampleref}[1]{\hyperref[#1]{{Example~\ref*{#1}}}}
\newcommand{\hypref}[1]{\hyperref[#1]{{Hypothesis~\ref*{#1}}}}
\newcommand{\secref}[1]{\hyperref[#1]{{Section~\ref*{#1}}}}
\newcommand{\chapref}[1]{\hyperref[#1]{{Chapter~\ref*{#1}}}}
\newcommand{\apref}[1]{\hyperref[#1]{{Appendix~\ref*{#1}}}}
\newcommand\rank{\mbox{\tt {rank}}\xspace}
\newcommand\prank{\mbox{\tt {rank}$_{\tt psd}$}\xspace}
\newcommand\alice{\mbox{\sf Alice}\xspace}
\newcommand\bob{\mbox{\sf Bob}\xspace}
\newcommand\pr{\mbox{\bf Pr}}
\newcommand\av{\mbox{\bf{\bf E}}}
\newcommand{\pabxy}{(p(ab|xy))}
\newcommand{\calQ}{\mathcal{Q}}

\newcommand{\comment}[1]{{}}
\newcommand{\blue}[1]{\textcolor{blue}{#1}}
\newcommand{\red}[1]{\textcolor{red}{#1}}
\newcommand{\green}[1]{\textcolor{green}{#1}}
\newcommand{\cyan}[1]{\textcolor{cyan}{#1}}
\newcommand{\magenta}[1]{\textcolor{magenta}{#1}}

\title{\vspace{-1cm} Quantifying Multipartite Quantum Entanglement in a Semi-Device-Independent Manner}

\author{Lijinzhi Lin$^{}$}
\author{Zhaohui Wei$^{}$}
\affiliation{$^{}$Center for Quantum
Information, Institute for Interdisciplinary
Information Sciences, Tsinghua University, Beijing 100084, China}

\begin{abstract}
	We propose two semi-device-independent approaches that are able to quantify unknown multipartite quantum entanglement experimentally, where the only information that has to be known beforehand is quantum dimension, and the concept that plays a key role is nondegenerate Bell inequalities. Specifically, using the nondegeneracy of multipartite Bell inequalities, we obtain useful information on the purity of target quantum state. Combined with an estimate of the maximal overlap between the target state and pure product states and a continuous property of the geometric measure of entanglement we shall prove, the information on purity allows us to give a lower bound for this entanglement measure. In addition, we show that a different combination of the above results also converts to a lower bound for the relative entropy of entanglement. As a demonstration, we apply our approach on 5-partite qubit systems with the MABK inequality, and show that useful lower bounds for the geometric measure of entanglement can be obtained if the Bell value is larger than 3.60, and those for the relative entropy of entanglement can be given if the Bell value is larger than 3.80, where the Tsirelson bound is 4.
\end{abstract}

\maketitle

\section{Introduction}

Quantum entanglement plays a fundamental role in quantum physics and quantum information, where it often serves as the key factor in physical effects or key resource in information processing tasks \cite{Horodecki09,NC00}. Therefore, how to certify the existence of quantum entanglement and even quantify it in physical experiments are two important problems. However, due to the imperfection of quantum operations and inevitable quantum noise, fulfilling these two tasks \emph{reliably} is extremely challenging. As a result, though some methods like entanglement witnesses have been applied widely in quantum laboratories~\cite{GT09}, they usually depend heavily on accurate knowledge on involved quantum systems, and possibly give incorrect results when it is not fully available~\cite{RFB+12}. Meanwhile, some other methods, like quantum tomography, consume too much resources, making it hard to apply them on large systems~\cite{CN97,PCZ97}.

To overcome these difficulties, a promising idea is to design protocols for these tasks in such a way that beforehand assumption needed on involved quantum systems, particularly on the precisions of quantum devices or quantum operations, is as little as possible, which allows us to draw reliable conclusions on quantum entanglement that we are interested. Following this idea, various \emph{device-independent} approaches have been proposed to tackle the problem of characterizing unknown quantum entanglement~\cite{Ekert91,BHK05,AGM06,PR92,MY98}. The key idea of these approaches is that the judgements are only based on quantum nonlocality that we can observe in quantum laboratories reliably, where one has to build nontrivial relations between quantum nonlocality and the aspects of quantum entanglement that we want to know. Indeed, a lot of interesting results of this kind have been reported or even demonstrated to certify the existence of genuine multipartite entanglement~\cite{CGP+02,BGLP11,PV11,MRMT16,BCWA17,TARGB18,ZDBS19}.

If only focusing on the issue of \emph{quantifying} unknown quantum entanglement experimentally, a lot of results have also been reported under the idea of device-independence~\cite{MBL+13,SHR17,BLM+09,Jed16,WL19}. For example, inspired by the Navascues-Pironio-Acin (NPA) method~\cite{NPA07}, in Ref.\cite{BLM+09} a device-independent approach to quantify the negativity, a measure of entanglement~\cite{ZHSL98}, was provided. In Ref.\cite{SHR17}, based on the idea of semiquantum nonlocal games~\cite{Buscemi12}, an approach that quantifies negative-partial-transposition entanglement was reported, where one does not have to put any trust onto measurement devices. In Ref.\cite{Jed16}, a new method with excellent performance was proposed to characterize the quantitative relation between entanglement measure and Clauser-Horne-Shimony-Holt inequality violations.

Particularly, in Ref.\cite{WL19} another general approach that is able to provide analytic results on entanglement measures, like the entanglement of distillation and the entanglement of formation, was proposed. Basically, this is a semi-device-independent approach, where the only assumption that we have to make beforehand is quantum dimension, and the key idea of this approach is introducing the concept of nondegenerate Bell inequalities, which plays a crucial role in providing nontrivial information on the purities of target quantum states. As a result, the purity information allows us to quantify the target entanglement by lower bounding coherent information, which is known to be a lower bound for the entanglement measures that we are interested~\cite{COF11}. However, an apparent drawback of the approach in Ref.\cite{WL19} is that it only works for bipartite entanglement.

For multipartite quantum entanglement, it has been known that its mathematical characterization, especially quantification, is a notoriously hard problem. However, it turns out that the geometric measure of entanglement (GME) and the relative entropy of entanglement (REE) are two quite successful measures for multipartite entanglement~\cite{BH01,WG03,VPRK97,VP98}. In this paper, we propose two theoretical approaches to quantify the two above measures of unknown multipartite quantum states in a semi-device-independent manner. The concept of nondegenerate Bell inequalities is essential to these approaches. Indeed, combined with the purity information provided by applying nondegenerate Bell inequality onto experimental statistics data, we manage to lower bound the GME by proving a continuous property of this entanglement measure. Furthermore, with the help of the purity information, we show that the REE can also be quantified by estimating the maximal overlap between the target state and pure product states. To achieve these tasks, we need to certify the nondegeneracy of multipartite Bell inequalities.

As a demonstration of our approaches, we show that the Mermin-Ardehali-Belinskii-Klyshko (MABK) inequality~\cite{Mermin90,Ardehali92,BK93} is nondegenerate for qubit systems, and then we demonstrate that nontrivial lower bounds on the GME and the REE of multipartite quantum entanglement can be obtained when the violation of the MABK inequality is sufficient, where it can be seen that the approaches have decent performance.

\section{Nondegenerate Bell inequalities}

Bell inequalities are crucial tools in the current paper, and in history they played a key role in the development of quantum mechanics~\cite{Bell64}. In a so-called $n$-partite Bell settings, $n$ space-separated parties share a physical system. Each party, say $i\in[n]\equiv\{1,2,...,n\}$, has a set of measurement devices labelled by a finite set $X_i$, and the corresponding set of possible measurement outcomes are labelled by a finite set $A_i$. Without communications, all parties choose random measurement devices from their own $X_i$ to measure their subsystems respectively, and record the outcomes. By repeating the whole process for sufficient times, they find out the joint probability distribution of outcomes for any given choices of measurement devices, denoted $p(a_1a_2...a_n|x_1x_2...x_n)$, where $x_i\in X_i$ and $a_i\in A_i$.

For simplicity, we call the above joint probability distribution a \emph{quantum correlation}, and write it as $p(\vec{a}|\vec{x})$ (or just $p$ if the context is clear), where $\vec{a}=(a_1a_2...a_n)$ and $\vec{x}=(x_1x_2...x_n)$. Then a (linear) Bell inequality is a relation that $p(\vec{a}|\vec{x})$ must obey if the system is classical, and it can be expressed as
\begin{equation}
I(p)=\sum_{\vec{a},\vec{x}}c_{\vec{x}}^{\vec{a}}p(\vec{a}|\vec{x})\leq C_l,
\end{equation}
where for any $\vec{x}$ and $\vec{a}$, $c_{\vec{x}}^{\vec{a}}$ is a real number.

However, a remarkable fact on quantum mechanics is that, if the shared physical system is quantum, Bell inequalities can be violated. Suppose the shared quantum state is $\rho$, then according to quantum mechanics $p(\vec{a}|\vec{x})$ can be written as
\begin{equation}
p(\vec{a}|\vec{x})=\Tr\left(\left(\bigotimes\limits_{i=1}^{n}M_{x_i}^{a_i}\right)\rho\right),
\end{equation}
where for any $i$ and $x_i$, $M_{x_i}^{a_i}$ is the measurement operators with outcome $a_i$ for the measurement with label $x_i$ performed by the $i$-th party. For convenience of later discussions, we let $I(\rho,M_{x_1}^{a_1},...,M_{x_n}^{a_n})$ be the Bell value achieved by $\rho$ and $M_{x_i}^{a_i}$. Then as mentioned above, if we let
\begin{equation}
C_q\equiv \max I(\rho,M_{x_1}^{a_1},...,M_{x_n}^{a_n}),
\end{equation}
where the maximum is taken over all possibilities of $\rho$ and $M_{x_1}^{a_1},...,M_{x_n}^{a_n}$, then it is possible that $C_q>C_l$, indicating that quantum systems are able to produce stronger correlations than classical ones.

In the joint quantum system, suppose the dimensions of the subsystems are $d_1,d_2,...,d_n$ respectively, then we call the vector $\vec{d}\equiv(d_1d_2...d_n)$ the \emph{dimension vector} of the joint system. In this paper, we are interested in the maximal value of $I(\rho,M_{x_1}^{a_1},...,M_{x_n}^{a_n})$ for fixed dimension vector $\vec{d}$. Similar with $C_q$, we denote it as $C_q(\vec{d})$.

%In this paper, in order to demonstrate our approach to quantify multipartite entanglement, we will use the MABK inequality as an example~\cite{Mermin90,Ardehali92,BK93}. For an $n$-partite quantum system, we let $X_i$ and $A_i$ be binary sets, then the corresponding MABK inequality reads
%\begin{widetext}
%\begin{align*}
%	\nonumber I_{\mathrm{MABK},n}(p)&=  {\left(\frac{1}{\sqrt{2}}\right)^{n-1}}\sum_{(x_1,\cdots,x_n)\in\{0,1\}^n}\sum_{(a_1,\cdots,a_n)\in\{0,1\}^n} \\
%	& \ \ \ \ \ \sin\left({\frac{(3-n)\pi}{4}}+\left(\sum_{i=1}^{n}(x_i+{2}a_i)\right)\frac{\pi}{2}\right)p(a_1\cdots a_n|x_1\cdots x_n)\\
%	& \leq {1}.
%\end{align*}
%\end{widetext}

The concept of nondegenerate Bell inequalities was proposed when studying bipartite quantum systems~\cite{WL19}. As we will see later, it can also be applied in the multipartite case and plays a key role in entanglement measure quantification.

Suppose $I\leq C_l$ is a Bell inequality for an $n$-partite quantum system with dimension $d_1\times d_2\times...\times d_n$. We say it is \emph{nondegenerate} on dimension vector $\vec{d}=(d_1...d_n)$, if there exist two real number $0\leq\epsilon_1<\epsilon_2\leq C_q(\vec{d})$, such that for any two quantum states of this system, $\ket{\alpha}$ and $\ket{\beta}$ with $\braket{\alpha}{\beta}=0$, and any quantum measurement sets $M_{x_1}^{a_1},...,M_{x_n}^{a_n}$, the relation that
\begin{equation*}
I(\ketbra{\alpha}{\alpha},M_{x_1}^{a_1},...,M_{x_n}^{a_n})\geq C_q(\vec{d})-\epsilon_1
\end{equation*}
always implies that
\begin{equation*}
I(\ketbra{\beta}{\beta},M_{x_1}^{a_1},...,M_{x_n}^{a_n})\leq C_q(\vec{d})-\epsilon_2.
\end{equation*}
Roughly speaking, if $I$ is a nondegenerate Bell inequality on dimension vector $\vec{d}$, then for any two orthogonal quantum states, at most one of them is able to violate $I$ remarkably using the same measurements.

We further let $M=\sum_{\vec{a},\vec{x}}c_{\vec{x}}^{\vec{a}}\left(\bigotimes\limits_{i=1}^{n}M_{x_i}^{a_i}\right)$, then it can be seen that $M$ is a Hermitian operator. And for any $\rho$ with dimension vector $\vec{d}$, it holds that $I(\rho,M_{x_1}^{a_1},...,M_{x_n}^{a_n})=\Tr(\rho M)$. Suppose $\lambda_1(M)\geq\cdots\geq\lambda_r(M)$ are the eigenvalues of $M$, where $r=d_1\times\cdots\times d_n$. For any integer $t$ with $1\leq t\leq r$, let
\begin{equation*}
C(I,\vec{d},t)\equiv\max \sum_{k=1}^t\lambda_{k}(M),
\end{equation*}
where the maximum is taken over all possible local quantum measurements. Then we immediately have that $C_q(\vec{d})=C(I,\vec{d},1)$. Furthermore, an important fact that allows us to certify the nondegeneracy of Bell inequalities is that, for any multipartite Bell inequality $I$ and any dimension vector $\vec{d}$, $I$ is nondegenerate if and only if $C(I,\vec{d},2)<2C(I,\vec{d},1)$, and when $I$ is nondegenerate, the parameters can be chosen by the relations $\epsilon_1<C(I,\vec{d},1)-\frac{1}{2}C(I,\vec{d},2)$ and $\epsilon_1+\epsilon_2=2C(I,\vec{d},1)-C(I,\vec{d},2)$~\cite{WL19}.

To illustrate the existence of nondegenerate multipartite Bell expressions, we consider the MABK expression over qubits~\cite{Mermin90,Ardehali92,BK93}. In fact, the nondegeneracy property of this inequality has been observed in Ref.\cite{SG01}, where it was proved that the first two eigenvalues of the Bell operator satisfy $\lambda_1^2(M)+\lambda_2^2(M)\leq 2^{n-1}$. This implies that if $\lambda_1(M)> 2^{n/2-1}$, we have
\begin{equation*}
\lambda_2(M)\leq \sqrt{2^{n-1}-\lambda_1^2(M)}<\lambda_1(M),
\end{equation*}
which indicates $C(I,\vec{d},2)<2C(I,\vec{d},1)$. Meanwhile, it is known that the maximal value that $\lambda_1(M)$ can achieve is $2^{n/2-1/2}$, where the corresponding state can be the $n$-qubit Greenberger-Horne-Zeilinger (GHZ) state $\ket{\text{GHZ}}=\frac{1}{\sqrt{2}}(\ket{0}^{\otimes n}+\ket{1}^{\otimes n})$~\cite{Mermin90,Ardehali92,BK93}. Therefore, the MABK expression is nondegenerate over qubits.

\section{Quantifying the geometric measure of entanglement}

The geometric measure of entanglement is a well-known measure for multipartite quantum entanglement~\cite{BH01,WG03}. Suppose $\ket{\psi}$ is a pure state of a joint system composed by $n$ subsystems. Define $G(\ket{\psi})$ to be the maximal overlap between $\ket{\psi}$ and a product pure state, that is to say,
\begin{equation}
G(\ket{\psi})=\sup_{\ket{\phi}\in\text{sep}_n}|\braket{\psi}{\phi}|,
\end{equation}
where $\text{sep}_n$ is the set of $n$-partite product pure states. Then for $\ket{\psi}$, its geometric measure of entanglement is defined to be
\begin{equation}
E_G(\ket{\psi})\equiv 1-G(\ket{\psi})^2.
\end{equation}
For a mixed state $\rho$ of this joint system, the geometric measure can be defined by convex roof construction, which is
\begin{equation}
E_G(\rho)\equiv \min_{\rho=\sum_ip_i\ketbra{\psi_i}{\psi_i}}\sum_{i}p_iE_G(\ket{\psi_i}).
\end{equation}

The GME has many nontrivial applications in quantum physics and quantum information, for example quantifying the difficulty of multipartite state discrimination under local operations and classical communications (LOCC)~\cite{MMV07}, constructing entanglement witness~\cite{WG03,HMM+08}, characterizing ground states of condensed matter systems and detecting phase transitions~\cite{ODV08,Orus08}, and so on. Therefore, it will be very nice if we can quantify the GME reliably in quantum laboratories. We now show how the concept of nondegenerate Bell inequalities allows us to achieve this, and the approach is composed by three steps as below.

\textbf{Step 1 }Suppose $\rho$ is the global state that produces the quantum correlation $p(\vec{a}|\vec{x})$.
Let the underlying measurements be $M_{x_1}^{a_1},...,M_{x_n}^{a_n}$; that is, $p(\vec{a}|\vec{x})= \mathrm{tr}\left(\left(\bigotimes_{i=1}^{n}M_{x_i}^{a_i}\right)\rho\right)$. Now, since a crucial component in the definition of GME is the maximum overlap
\begin{align*}
	G(\ket{\psi})=\max\limits_{\ket{\phi}\in\text{sep}_n}|\braket{\phi}{\psi}|=\max\limits_{\ket{\phi}\in\text{sep}_n}F(\ket{\phi},\ket{\psi})
\end{align*}
where $F$ is the fidelity, we wish to quantify the related fidelity
\begin{align*}
	\max\limits_{\ket{\phi}\in\text{sep}_n}F(\ket{\phi}\bra{\phi},\rho)=\max\limits_{\ket{\phi}\in\text{sep}_n}\sqrt{\bra{\phi}\rho\ket{\phi}}
\end{align*}
in a fully device-independent manner, where $\text{sep}_n$ is the set of product pure states.

Suppose $\ket{\phi}\in\text{sep}_n$ is the state that maximizes $F(\ket{\phi}\bra{\phi},\rho)$.
Let $q^*(\vec{a}|\vec{x})$ be the correlation produced by $\ket{\phi}$ upon measurements $M_{x_1}^{a_1},...,M_{x_n}^{a_n}$.
Since $\ket{\phi}$ is a product pure state, the correlation $q^*$ is a product correlation;
that is, there exists probability distributions $q^*_i(a_i|x_i)$ such that $q^*(\vec{a}|\vec{x})=\prod_{i=1}^{n}q^*_i(a_i|x_i)$.
When $\rho$ and $\ket{\phi}$ are measured, the fidelity between them should increase~\cite{NC00};
that is, for any $\vec{x}$ the resulting probability distribution $p_{\vec{x}}\equiv p(\cdot|\vec{x})$ and $q^*_{\vec{x}}\equiv q^*(\cdot|\vec{x})$ satisfy
\begin{align*}
	\sum_{\vec{a}}\sqrt{q^*(\vec{a}|\vec{x})p(\vec{a}|\vec{x})}= F(q^*_{\vec{x}},p_{\vec{x}})\geq F(\ket{\phi}\bra{\phi},\rho),
\end{align*}
hence it holds that $\min\limits_{\vec{x}}F(q^*_{\vec{x}},p_{\vec{x}})\geq F(\ket{\phi}\bra{\phi},\rho)$. Since $q^*$ is a product correlation, we have
\begin{align*}
	\max\limits_{q}\min\limits_{\vec{x}}F(q_{\vec{x}},p_{\vec{x}})\geq F(\ket{\phi}\bra{\phi},\rho),
\end{align*}
where the outmost maximization is over product correlations $q$ and $q_{\vec{x}}\equiv q(\cdot|\vec{x})$.
By the max-min inequality, it holds that
\begin{align*}
	\min\limits_{\vec{x}}\max\limits_{q}F(q_{\vec{x}},p_{\vec{x}})\geq \max\limits_{q}\min\limits_{\vec{x}}F(q_{\vec{x}},p_{\vec{x}}),
\end{align*}
then we have
\begin{align*}
	\min\limits_{\vec{x}}\max\limits_{q}F(q_{\vec{x}},p_{\vec{x}})\geq F(\ket{\phi}\bra{\phi},\rho).
\end{align*}
Then by numerical calculations on the correlation data, we can get an upper bound on the fidelity between the target state and a pure product state, denoted as $\hat{F}$. For example, once $\vec{x}$ is fixed, the inner maximization can be computed using symmetric embedding~\cite{RV13} and the shifted higher-order power method (SHOPM) algorithm~\cite{KM11}, yielding a correct answer up to numerical precision with very high probability (see also Ref.\cite{HQZ16}).

\textbf{Step 2 }Since computing GME for a mixed state requires complicated optimization over ensembles, it would be ideal for the quantification of GME if $\rho$ is a pure state. Therefore, we wish to bound the purity of $\rho$, defined as $\Tr(\rho^2)$, from below, which is accomplished by the nondegeneracy property of Bell inequalities~~\cite{WL19}.

Let $\rho=\sum_i a_i\ket{\psi_i}\bra{\psi_i}$ be the spectral decomposition of $\rho$. Suppose $I$ is a nondegenerate Bell expression with parameters $\epsilon_1$ and $\epsilon_2$ satisfying $0\leq \epsilon_1<\epsilon_2$.
If $I(\rho,M_{x_1}^{a_1},...,M_{x_n}^{a_n})\geq C(I,\vec{d},1)-\epsilon_1$, then there is $i$ such that $I(\ket{\psi_i}\bra{\psi_i},M_{x_1}^{a_1},...,M_{x_n}^{a_n})\geq C(I,\vec{d},1)-\epsilon_1$.
Thus, by nondegeneracy of $I$, we have
\begin{align*}
	C(I,\vec{d},1)-\epsilon_1\leq & I(\rho,M_{x_1}^{a_1},...,M_{x_n}^{a_n})\\
       =&\sum_j a_j I(\ket{\psi_j}\bra{\psi_j},M_{x_1}^{a_1},...,M_{x_n}^{a_n}) \\
	\leq & a_i C(I,\vec{d},1)+(1-a_i)(C(I,\vec{d},1)-\epsilon_2).
\end{align*}
This implies that $a_i\geq 1-\epsilon_1/\epsilon_2$. Since the order of eigenstates in the spectral decomposition is arbitrary, for convenience we now relabel the index $i$ found above to $1$, then it holds that $a_1\geq 1-\epsilon_1/\epsilon_2$. This allows us to lower bound the purity.

\textbf{Step 3 }In the previous two steps, we obtained a lower bound for $a_1$ in the spectral decomposition of $\rho$
and an upper bound $\hat{F}$ for $F(\ket{\phi}\bra{\phi},\rho)=\sqrt{\bra{\phi}\rho\ket{\phi}}$ among all product pure states $\ket{\phi}$.
The following theorem shows that, if $\hat{F}\leq a_1$, then we can derive a lower bound for $E_{G}(\rho)$ by proving a continuous property of GME. The proof for this theorem can be seen in the appendix.
%Since GME is extended to mixed states via convex roof construction
%and the maximum overlap is continuous over pure states,
%it follows that GME is continuous.
%This allows us to obtain a lower bound for the GME of $\rho$ when $\hat{F}$ is sufficiently small and $a_1$ is sufficiently high, which is captured by the following proposition.
\begin{theorem*}
	\label{prop:cont}
	Suppose $\hat{F}\leq a_1$, then it holds that
\end{theorem*}

  \begin{widetext}
   \begin{eqnarray*}
    E_{G}(\rho)\geq \max_{c\in\left[\frac{\hat{F}}{\sqrt{a_1}},\sqrt{a_1}\right]}\frac{a_1-c^2}{1-c^2}\left(1-\left(\frac{\hat{F}}{\sqrt{a_1}}c+\sqrt{1-\frac{\hat{F}^2}{a_1}}\sqrt{1-c^2}\right)^2\right).
  \end{eqnarray*}
  \end{widetext}
In particular, if $\rho$ is a pure state, then it holds that $a_1=1$. In that case, the lower bound in Proposition \ref{prop:cont} reads
\begin{align}
	E_{G}(\ket{\psi_1}\bra{\psi_1})\geq 1-\hat{F}^2,
\end{align}
which agrees with the definiton of GME on pure states, indicating that our lower bound is tight in this case.

Therefore, combining all the above three steps together, we obtain a semi-device-independent approach to quantify the GME for unknown multipartite entanglement. %First, based on the statistics data of the Bell experiment, we lower bound the purity of the target state by using a proper nondegenerate Bell inequality, and then upper bound its maximal overlap with a product pure state. Second, according to the continuity of the GME, if the above two bounds are good enough, we can obtain a lower bound for the GME of the target state.
We now demonstrate that this approach indeed works well by quantifying the GME of an $n$-partite quantum system with the MABK inequality ($n=3,5$). Recall that we have known that this inequality is nondegenerate. At the same time, we would like to stress that in principle the approach can be applied on any multipartite quantum systems with known dimensions.

There exist many configurations that achieve the maximum violation to the MABK inequality, and it turns out that they are essentially equivalent~\cite{Jed17}. For example, one can let the state be
	\begin{align*}
		\ket{\Phi}=\frac{1}{\sqrt{2}}\left(\ket{0}^{\otimes n}+e^{\frac{2\pi i}{8}(n-1)}\ket{1}^{\otimes n}\right),
	\end{align*}
then measure the observables $\sigma_x$ and $\sigma_y$ on each qubit.
That is, for each site, we select
\begin{align*}
	M^{0}_{0}= & \ket{+}\bra{+}, \\
	M^{1}_{0}= & \ket{-}\bra{-}, \\
	M^{0}_{1}= & \ket{{+i}}\bra{+i}, \\
	M^{1}_{1}= & \ket{{-i}}\bra{-i},
\end{align*}
where $\ket{\pm}=1/\sqrt{2}(\ket{0}\pm\ket{1})$ and $\ket{\pm i}=1/\sqrt{2}(\ket{0}\pm i\ket{1})$.

\begin{figure}[htbp]
	\centering
	\includegraphics[width=3.5in]{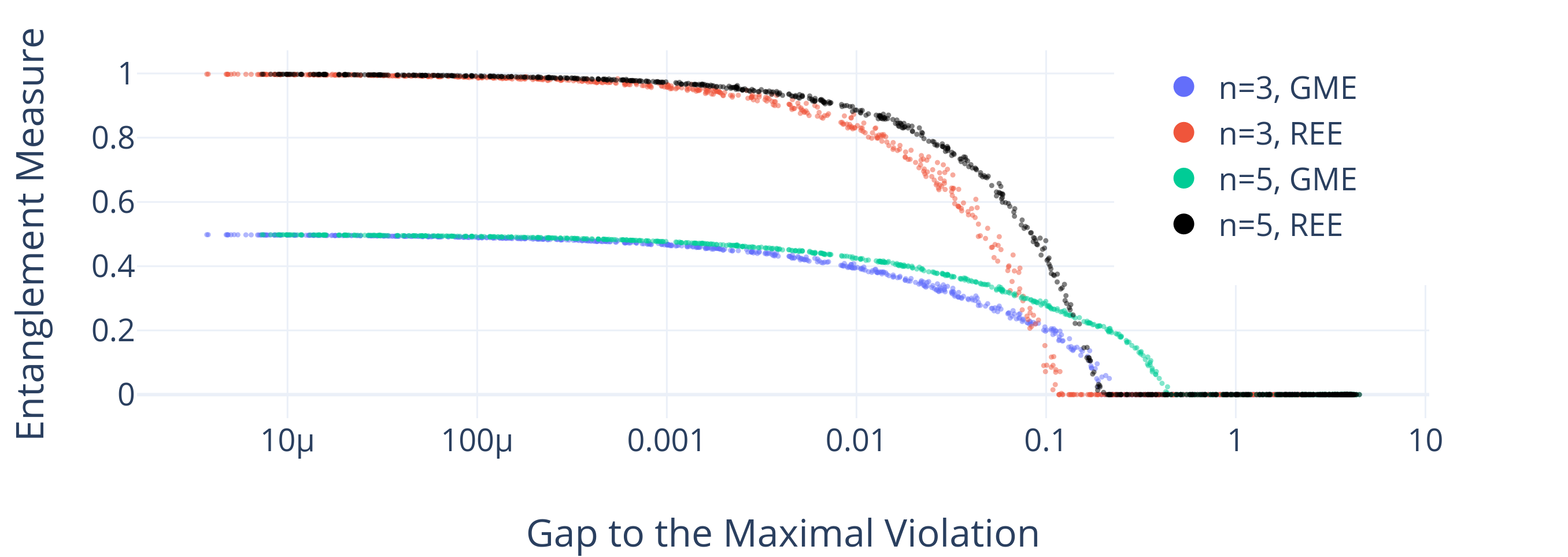}
	\caption{Lower bounds for the GME and the REE, where $n=3,5$. Note that the maximal Bell value is 2 ($n=3$) and 4 ($n=5$) respectively, and we focus on the gaps to the maximal Bell values.}
	\label{fig:example}
\end{figure}

To obtain physical statistic data of the Bell experiments, we perturb the state $\ket{\Phi}$ and the above optimal measurements, which produces a series of legitimate quantum correlations. We then apply our approach to each correlation. The result is shown in Fig \ref{fig:example}.

It turns out that when $n=3$, if the Bell expression value is more than $1.80$, our approach is able to provide nontrivial result on the GME. As a comparison, the Tsirelson bound for this case is $2$. Furthermore, when the violation approaches the maximum, our approach gives a tight result $0.5$, considering that the maximal violation is achieved by $\ket{\Phi}$.

Similarly, Fig \ref{fig:example} also illustrates the result of our approach on 5-partite qubit systems, where the same patterns with the case $n=3$ can be observed. Here, nontrivial GME lower bounds can be obtained when the Bell expression value is more than $3.60$, where the Tsirelson bound is $4$.

\section{Quantifying the relative entropy of entanglement}

Interestingly, Step 1 and Step 2 introduced above are already sufficient for us to lower bound the relative entropy of entanglement (REE) in a semi-device-independent manner.

The REE of $\rho$ is defined to be the minimal relative entropy of $\rho$ to the set of separable states, that is,
\begin{align}
	E_{R}(\rho)\equiv\min_{\sigma\in\mathcal{D}}S(\rho\Vert\sigma)=\min_{\sigma\in\mathcal{D}}\Tr(\rho\log_2\rho-\rho\log_2\sigma),
\end{align}
where $\mathcal{D}$ is the set of all separable states~\cite{VPRK97,VP98}. It turns out that the REE has many profound applications in quantum information theory.
For example, for bipartite quantum states, REE can lower bound the entanglement of formation and upper bound the entanglement of distillation~\cite{VP98}.
Therefore, quantifying the REE reliably in experiments is naturally very important and rewarding.

We now show that $E_{R}(\rho)$ has a close relation with the quantity $\hat{F}$ introduced above. In fact, it has been known that~\cite{Wei08}
\begin{align}
	E_{R}(\rho)\geq G(\rho)-S(\rho),
\end{align}
where $S(\rho)$ is the Von Neumann entropy and
\begin{align}
	G(\rho)\equiv-\log_2\left\{\max\limits_{\ket{\phi}\in\text{sep}_n}\bra{\phi}\rho\ket{\phi}\right\}.
\end{align}
Since $\max\limits_{\ket{\phi}\in\text{sep}_n}\bra{\phi}\rho\ket{\phi}\leq\hat{F}^2$, it holds that
\begin{align}\label{eq:ree}
	E_{R}(\rho)\geq -2\log_2(\hat{F})-S(\rho).
\end{align}
In the meantime, in Step 2 we get a lower bound for the purity of $\rho$ (in terms of $a_1$). Combining this fact and the approach introduced in Ref.\cite{SSY+17}, we can derive a upper bound for $S(\rho)$ (see \cite{WL19} for a complete demonstration). According to Eq.\eqref{eq:ree}, this implies that we are able to lower bound the REE and any other multipartite entanglement measures that are lower bounded by the REE (for example, the generalized robustness of entanglement~\cite{VT99,Steiner03,Wei08}).

Still using the MABK inequality and the samples of quantum correlations generated above, we test the performance of the second approach, and the result can also be seen in Fig \ref{fig:example}. Particularly, when $n=3$, our approach can give positive lower bound for the REE when the Bell value is larger than 1.88; when $n=5$, it can provide nontrivial results when the Bell value is larger than 3.80.

\section{Conclusion}

Based on the concept of nondegenerate Bell inequalities, we show that multipartite quantum entanglement can be quantified experimentally in a semi-device-independent way. The key information provided by this concept is on the purity of the target quantum systems. Based on this, by studying the mathematical properties of the geometric measure of entanglement and the relative entropy of entanglement, we can provide nontrivial lower bounds for these two well-known entanglement measures. Our approaches do not need any trust on the precision of the involved quantum devices except for their dimensions and have decent performance. We hope that these approaches would prove to be valuable for characterizing unknown multipartite states in future quantum experiements.

\begin{acknowledgments}
 We thank Yu Guo, Yongjian Han, Biheng Liu, and J\c{e}drzej Kaniewski for helpful comments on an earlier draft, and Huangjun Zhu for help discussions. L.L. and Z.W. are supported by the National Key R\&D Program of China, Grant No. 2018YFA0306703 and the start-up funds of Tsinghua University, Grant No. 53330100118. This work has been supported in part by the Zhongguancun Haihua Institute for Frontier Information Technology.
\end{acknowledgments}

\appendix
\section{The proof for the Theorem}
\begin{theorem*}
	Suppose $\hat{F}$ and $a_1$ are defined as in the text, and $\hat{F}\leq a_1$, then it holds that
\end{theorem*}

  \begin{widetext}
   \begin{eqnarray*}
    E_{G}(\rho)\geq \max_{c\in\left[\frac{\hat{F}}{\sqrt{a_1}},\sqrt{a_1}\right]}\frac{a_1-c^2}{1-c^2}\left(1-\left(\frac{\hat{F}}{\sqrt{a_1}}c+\sqrt{1-\frac{\hat{F}^2}{a_1}}\sqrt{1-c^2}\right)^2\right).
  \end{eqnarray*}
  \end{widetext}

\begin{proof}
	Suppose $\rho=\sum_j \tilde{a}_j\ket{\tilde{\psi}_j}\bra{\tilde{\psi}_j}$ is an ensemble of $\rho$ that obtains the GME of $\rho$.
	Let $c$ be a real number in the interval $[\hat{F}/\sqrt{a_1},\sqrt{a_1}]$.
	Consider the sets of indices
	\begin{align*}
		J_1= & \{j:{|\braket{\psi_1}{\tilde{\psi}_j}|}\geq c\}, \\
		J_2= & \{j:{|\braket{\psi_1}{\tilde{\psi}_j}|}<c\},
	\end{align*}
	which form a partition of the set of all indices $j$.
	Intuitively, the set $J_1$ consists of components with high fidelity with $\ket{\psi_1}$.
	Let $\mu=\sum_{j\in J_1}\tilde{a}_j$.
	We have
	\begin{align*}
		a_1= & \braket{\psi_1}{\rho|\psi_1} \\
		= & \sum_j \tilde{a}_j|\braket{\psi_1}{\tilde{\psi}_j}|^2 \\
		= & \sum_{j\in J_1} \tilde{a}_j|\braket{\psi_1}{\tilde{\psi}_j}|^2+\sum_{j\in J_2} \tilde{a}_j|\braket{\psi_1}{\tilde{\psi}_j}|^2 \\
		\leq & \mu+(1-\mu)c^2,
	\end{align*}
	thus
	\begin{align*}
		\mu\geq\frac{a_1-c^2}{1-c^2}\geq 0,
	\end{align*}
	which is lower bound for the sum of weights of components whose indices belong to $J_1$.
	Note that $\mu\to 1$ when $a_1\to 1$ if $c<\sqrt{a_1}$, and $\mu=1$ if $a_1=c=1$.
	By the definition of $\hat{F}$, for any product pure state $\ket{\phi}$, we have
	\begin{align*}
		\hat{F}^2\geq\braket{\phi}{\rho|\phi}=\sum_i a_i|\braket{\phi}{\psi_i}|^2\geq a_1|\braket{\phi}{\psi_1}|^2,
	\end{align*}
	thus
	\begin{align*}
		|\braket{\phi}{\psi_1}|\leq \frac{\hat{F}}{\sqrt{a_1}}.
	\end{align*}
	On the other hand, there are states $\{\phi_j\}$ such that
	\begin{align*}
		E_{G}(\rho)= & 1-\sum_j \tilde{a}_j|\braket{\phi_j}{\tilde{\psi}_j}|^2.
	\end{align*}
	By the triangle inequality of fidelity, for every $j\in J_1$, we have
	\begin{align*}
		\arccos|\braket{\phi_j}{\tilde{\psi}_j}|\geq & \arccos|\braket{\phi_j}{\psi_1}|-\arccos|\braket{\psi_1}{\tilde{\psi}_j}| \\
		\geq & \arccos\left(\frac{\hat{F}}{\sqrt{a_1}}\right)-\arccos(c).
	\end{align*}
	As $\hat{F}/\sqrt{a_1}\leq c$, the inequality above implies
	\begin{align*}
		|\braket{\phi_j}{\tilde{\psi}_j}|\leq \frac{\hat{F}}{\sqrt{a_1}}c+\sqrt{1-\frac{\hat{F}^2}{a_1}}\sqrt{1-c^2}.
	\end{align*}
	For $j\in J_2$, we upper-bound the overlap via $|\braket{\phi_j}{\tilde{\psi}_j}|\leq 1$,
	thereby obtaining a lower bound for the GME of $\rho$ as
    \begin{eqnarray*}
    	E_{G}(\rho)\geq & 1-\mu\left(\frac{\hat{F}}{\sqrt{a_1}}c+\sqrt{1-\frac{\hat{F}^2}{a_1}}\sqrt{1-c^2}\right)^2-(1-\mu) \\
    	= & \mu\left(1-\left(\frac{\hat{F}}{\sqrt{a_1}}c+\sqrt{1-\frac{\hat{F}^2}{a_1}}\sqrt{1-c^2}\right)^2\right) \\
    	\geq & \frac{a_1-c^2}{1-c^2}\left(1-\left(\frac{{\hat{F}}}{\sqrt{a_1}}c+\sqrt{1-\frac{\hat{F}^2}{a_1}}\sqrt{1-c^2}\right)^2\right).
    \end{eqnarray*}
   Note that the above relation holds for any $c\in[\hat{F}/\sqrt{a_1},\sqrt{a_1}]$, which concludes the proof.
\end{proof}

\end{document}